\documentclass[amsmath,amssymb,aps,prl,twocolumn]{revtex4-2}
\pdfoutput=1
\usepackage[utf8]{inputenc}
\usepackage{epsfig,subfigure,amsmath,float,xcolor,ulem}
\usepackage[colorlinks]{hyperref}
\hypersetup{colorlinks=true,linkcolor=blue,
filecolor=blue,urlcolor=blue,citecolor=blue}
\newcommand\beq{\begin{equation}}
\newcommand\eeq{\end{equation}}
\newcommand\bes{\begin{subequations}}
\newcommand\ees{\end{subequations}}
\newcommand\bea{\begin{eqnarray}}
\newcommand\eea{\end{eqnarray}}
\newcommand\non{\nonumber}

\newcommand\ig{\includegraphics}

\newcommand\al{\alpha}

\newcommand\ga{\gamma}

\newcommand\De{\Delta}
\newcommand\eps{\epsilon}
\newcommand\si{\sigma}

\newcommand\la{\langle}
\newcommand\ra{\rangle}
\newcommand\Om{\Omega}
\newcommand\om{\omega}

\newcommand\vk{{\bf k}}

\begin{document}
\title{Optical conductivity of a topological system driven using a realistic pulse}
\author{Ranjani Seshadri${}^{1,2}$}
\email{ranjanis@post.bgu.ac.il}
\author{T. Pereg-Barnea${}^{2}$}
\email{tamipb@physics.mcgill.ca}
\affiliation{$^1$Department of Physics, Ben-Gurion University of the Negev,
Beer-Sheva 84105, Israel}
\affiliation{{$^2$Department of Physics, McGill University, Montr\'eal,Qu\'ebec H3A 2T8, Canada}}
\begin{abstract}
The effect of a time-periodic perturbation, such as radiation, on a system otherwise at equilibrium has been
studied in the context of Floquet theory with stationary states replaced by Floquet states and the energy
replaced by quasienergy. These quasienergy bands in general differ from the energy bands in their dispersion
and, especially in the presence of spin-orbit coupling, in their states.  This may, in some cases, alter the
topology when the quasienergy bands exhibit different topological invariants than their stationary counterparts.
In this work, motivated by advances in pump-probe techniques, we consider the optical response of driven
topological systems when the drive is not purely periodic but is instead multiplied by a pulse shape/envelope
function. We use real time-evolved states to calculate the optical conductivity and compare it to the response
calculated using Floquet theory. We find that the conductivity bears a memory of the initial equilibrium
state even when the pump is turned on slowly and the measurement is taken well after the ramp. The response of
the time-evolved system is interpreted as coming from Floquet bands whose population has been determined by their
overlap with the initial equilibrium state.  In particular, at band inversion points in the Brillouin zone the
population of the Floquet bands is inverted as well. 
\end{abstract}

\date{\today}
\maketitle

{\it Introduction-}
The theoretical prediction and experimental realization of topological insulators (TIs)  
\cite{Moore1,Moore2,fu1,hasan,qi,BHZ1,hseih,konig,roth} has been one of the greatest developments in
condensed matter physics in the last decade.  Not only do topological insulators represent a paradigm
shift in condensed matter physics, they are also predicted to have a variety of applications
\cite{wang19, Legg2022, ali, Yao2012, yue16}.

While spin-orbit coupling is a key ingredient, it need not always lead to non-trivial topology as band
inversion may not always occur or the presence of a Fermi surface may not be avoided. It has therefore
been proposed to use a time-periodic perturbation in order to control the topology
\cite{lind,holt,kun,cal,top1,top2,top18a,top18,kit1,ten,titum,freg,foa,saha,sesh22,sesh23,gu,harp20,
Rudner2020,lin13,dora2012, thak2013, thak2014, kat2013}. 
When a time-periodic perturbation is added to a Hamiltonian the system is no longer invariant under an
arbitrary translation in time. However a reduced discrete time-translation symmetry still exists. This
allows finding solutions to the time-dependent Schr\"{o}dinger equation using Floquet theory. 
These solutions or Floquet states are eigenstates of the time evolution operator over a single drive
cycle. In other words, Floquet states are periodic up to a phase which is interpreted as $-\epsilon T$
where $\epsilon$ is the quasienergy and $T$ is the drive period. %{The quasienergy spectrum need not
%resemble the equilibrium dispersion and the Floquet states (spinor) may differ from the stationary
%states of the undriven system}.
The quasienergies and Floquet states (say, spinors) in general differ from the equilibrium energies and
eigenstates respectively. Therefore, along with other properties, topological invariants can change as a
result of irradiation leading to driven topological phase transitions. For example, graphene may be driven
into a topological phase \cite{gu} where gaps appear in the spectrum.  In case of Weyl semimetals, Weyl
nodes may split into Dirac points or gap out and give rise to Chern bands \cite{hubener} while spin-orbit
coupled insulators are predicted to become topological upon diving \cite{lind}.

While there are several theoretical predictions, the experimental realization of such Floquet-driven
topological phase transitions seems to be challenging.  Notably, analogue photonic systems
were the first to realize some of these predictions \cite{rec} and recently driven graphene has
shown signs of topology \cite{mciv} while the general idea of Floquet bands has
been demonstrated by time-resolved ARPES \cite{far1,wang}.

Several obstacles occur while trying to realize such Floquet-driven topological transitions. These include
sample heating, damping and disorder. But perhaps the most elementary deviation from the pure Floquet drive
is the unavoidable pulse shape. The drive can not be turned on at time $t=-\infty$ and therefore the state
of the system is always connected to that of the equilibrium state. One might expect that at long times
after the turning on of the drive, the state  will resemble a Floquet state. However, as will be shown
below, the notion of adiabaticity does not hold at relevant drive frequencies. In particular,
as will be discussed here, the system does not forget its initial conditions. While Floquet states may
be a good approximation for the single particle states at long time after the perturbation has
been turned on, their population is highly dependent on the initial state.  Therefore, one should not
expect completely filled or empty Floquet bands at low temperature meaning that the full potential of
topological invariance may not be realized. Moreover, it seems that relaxation effects do not
necessarily lead to the desired population as the quasienergy is periodic and energy may not
necessarily relax to one band \cite{deh3}. Similarly, when connecting a Floquet spin-Hall insulator to
leads, one can not measure quantized conductivity due to mismatch between the equilibrium states of the
leads and the driven system \cite{Aaron1,Aaron2}. 

The task at hand is therefore to accurately describe a system driven by a pulse of light whose width
is in the range of a few to many time periods, as appropriate for pump-probe techniques and 
understand the relation between the optical conductivity and Floquet band population.
%time-dependent problem without restricting to a single cycle of the drive. Studying the 
%behaviour 	of 	
%materials under such an aperiodic external perturbation is important in the context of 
%pump-probe 
%techniques which rely on ultra-short laser pulses. Owing to the rapid developments in the 
%field of 	
%ultra-fast lasers, this has emerged as a very useful tool to experimentally study quantum	
%matter. 
%Here an ultra-short laser pulse i.e. the `pump' is used to drive the sample out of 	
%equilibrium. The 
%non-equilibrium state thus generated is studied using a relatively weaker beam 	or 	the 
%`probe'. 
%Studying the response of a sample as a function of the time difference between the pump 
%and probe 
%gives us insights into the dynamics of charge carriers in the	sample.
In this work we are interested in the physics of such pump-probe measurements and how topological 
systems respond to a perturbation that breaks time periodicity. We look at the  behavior of the
Bernevig-Hughes-Zhang (BHZ) model of a two-dimensional TI \cite{BHZ1} in presence of a perturbation
in the form of a short pulse and compare it with the response of an exactly periodic (Floquet) drive.

%electrical conductance \cite{deh1, deh2, kum1}- both longitudinal and transverse components -  
%in the 

% We begin with a brief overview of the Bernevig-Hughes-Zhang (BHZ) model of a 
% two-dimensional TI in \ref{sec:BHZ} and describe the effect of a periodic perturbation on the
% energy spectrum. This is followed by an overview of Linear response theory in Sec. \ref{sec:linresp} 
% where we derive expressions for the frequency dependence of conductance. We then use 
% these in Sec. \ref{sec:top} to study the response of the BHZ system in the topological phase 
% to a non-periodic drive and compare it with the equilibrium and Floquet behaviour.

{\it Driven BHZ Model- }
The equilibrium Hamiltonian in momentum-space is written as
\beq
H_0(\vk) = {\bf d(\vk)}\cdot{\boldsymbol{\si}}  \label{eq:BHZHam}
\eeq
with ${\bf d}(\vk)=(A \sin k_x, A \sin k_y, M - 2B(2-\cos k_x - \cos k_y))$
and ${\boldsymbol{\si}} = (\si^x,\si^y,\si^z)$ are the $2\times2$ Pauli matrices. The mass $M$ and
hopping amplitude $B$ are expressed in units of the spin-orbit coupling strength $A$. The spectrum,
in general, is insulating in the bulk with a finite band gap. We work in a parameter regime ($M=0.2A$
and $B=0.2A$) where the system is topological. The equilibrium Chern numbers are
calculated numerically as explained in the Supplemental material \cite{supp} following the method prescribed
in Ref \cite{fukui} and are found to be $C^{Eq}_\pm = \pm 1$ for the top and bottom band respectively.
% \bea
% H_0(\vk)&=&(M-2B(2-\cos{k_x}-\cos{k_x}))\si^z \non \\
% &+& A(\sin{k_x}\si^x +\sin{k_y}~\si^y).\label{eq:Eq_BHZ}
% \eea

A time-dependence $H(t) = H_0+V(t) \si^z$ effectively makes the mass time-dependent.
In the ideal case, $V(t)$ is perfectly periodic with a frequency $\Om$. However, in
reality, this perfect periodicity cannot exist forever and is instead approximated by
realistic cases of a slow quench or a Gaussian pulse,
\begin{small}
\bea
V(t) = \begin{cases} V_{{Floq}}(t) = V_0 \sin(\Om t) \\
V_{{Pump}}(t) = V_0 \sin(\Om t) e^{-\frac{t^2}{2\De^2}}  \\
V_{{Quench}}(t) = V_0 \sin(\Om t) \frac{1+\tanh(\beta t)}{2}
\end{cases} 
\label{eq:vpump}
\eea
\end{small}
where $V_{{Floq}},V_{{Pump}}$ and $V_{{Quench}}$ have different envelope functions - constant, Gaussian
and a smooth ramp respectively.  Here $\beta$ and $\De$ are the rate of the quench and the width of
the pulse, respectively. $V_0$ ($=0.35 A$) is the peak amplitude of the perturbation
in all cases. The $V_{Floq}(t)$ with frequency $\Omega=2A$ drives the system to a topological phase
with Chern numbers $C^{Fl}_{\pm} = \pm2$.
\begin{figure}[htb]
\centering
\ig[width=8.64cm]{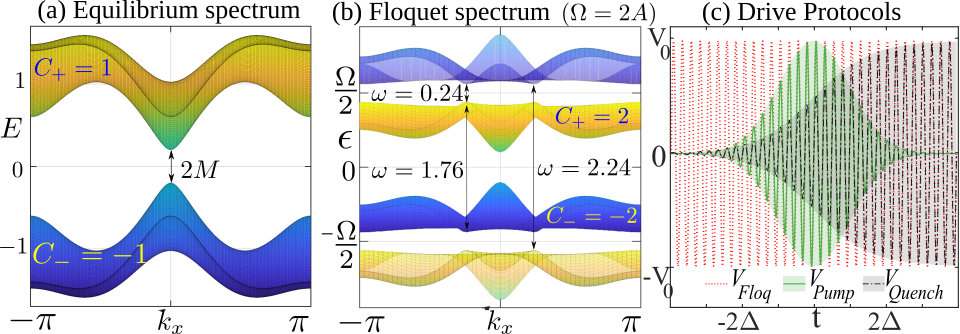}
\caption{(a) The gapped equilibrium spectrum with Chern number $\pm 1$.
(b) With a periodic drive of frequency $\Om = 2A$ the ideal Floquet spectrum is also gapped
with Chern numbers $\pm2$. 
%The primary Floquet zone
% $-\frac{\Omega}{2}\leq \epsilon \leq \frac{\Omega}{2}$ is shown in solid colors. 
The transitions
marked in (b) correspond to features marked as ${*}$s in Fig. \ref{fig:pk}.
(c) The drive protocols of in Eq. \eqref{eq:vpump} -
the ideal Floquet case (red dotted), the slow quench (black dot dash) and the Gaussian pump
(green solid) .}
\label{fig:pump}
\end{figure}

In the perfectly periodic case we employ Floquet theory to find the wave functions which are then used
to calculate the response functions, assuming that one of the Floquet bands is completely filled while
the other is completely empty. In the cases of quench and pump, the analysis requires an actual time
evolution over several drive cycles since the perfect periodicity is lost due to the envelope. The
response to a non-periodic drive (Gaussian or quench) is calculated using the time-evolved states 
starting from the equilibrium states of the lower band of the undriven system. We use these states
in the Kubo formula as described below.  

{\it Linear Response Theory - } According to Kubo formula the susceptibility,
which is in our case is the response of the driven system to a small probe field is given by,
\begin{small} 
\bea
\chi_{AB}(t,t')=\lim_{\eta\rightarrow 0^+} e^{\eta t'}\Big(&i\Theta(t-t')&Tr\Big\{g_0 [A^I(t'),B^I(t)]\Big\} \non \\
&+ \delta(t-t')&Tr \Big\{g_0 M^I(t)\Big\}\Big) \label{eq:lin1}
\eea
\end{small}
where $A^I$, $B^I$ and $M^I$ are operators in the interaction representation. The density matrix
$g_0$ determines the initial state of the system and $\eta >0$ is a small parameter used to smoothen
the response function. The Heaviside step-function $\Theta(t-t')$ ensures that causality is not violated.
The diamagnetic term $M^I(t)$ 
contributes only to the DC conductivity in the limit $\omega \rightarrow 0$. 

For computing electrical conductivity, both $A$ and $B$ are current operators.
As explained in the Supplemental material \cite{supp}, following reference
\cite{kum1}, Eq. \eqref{eq:lin1} becomes, 
\begin{small}
\bea
&\chi_{uv}(t,t')&= \lim_{\eta\rightarrow 0^+} e^{\eta t'} \sum_{\al \ga} g_{0\al}\Big[2 i \Theta(t-t') \non \\
&\times& \Big(\la\psi_\al(t')|J_u|\psi_\ga(t')\ra \la\psi_\ga(t) |J_v|\psi_\al(t)\ra  - {u \leftrightarrow v}\Big) \non \\
&+& \delta(t-t')\la\psi_\al(t)|M_{uv}|\psi_\al(t)\ra\Big]\label{eq:chitt}
\eea
\end{small}
% \onecolumngrid
% \begin{widetext}    
% \bea
% &\chi_{uv}(t,t')&= \lim_{\eta\rightarrow 0^+} e^{\eta t'} \sum_{\al \ga \vk} g_{0\al} (\vk) \Big[2 i \Theta(t-t') \non \\
% &\times&Im \Big(\la\psi_\al(\vk,t')|~H_u(\vk)~|\psi_\ga(\vk,t')\ra \la\psi_\ga(\vk, t) |~H_v(\vk)~|\psi_\al(\vk,t)\ra \Big) \non \\
% &+& \delta(t-t')\la\psi_\al(\vk,t)|H_{uv}(\vk)|\psi_\al(\vk,t)\ra\Big]\label{eq:chitt}
% \eea
% \end{widetext}
Here $|\psi_\al(t))\rangle$ is the state corresponding to band $\al$ at time $t$, $g_{0\al}$
gives the occupation of states at the initial time; the current operator $J_u = \partial_{k_u}H$ and the
inverse mass $M_{uv} = \partial_{k_u}\partial_{k_v}H$. The subscripts $u$ and $v$ are the in-plane directions
with $u=(\neq)v$ being longitudinal (transverse) conductivity. and the $\vk$-dependence has been skipped
for brevity. In the specific case of the model we have considered, the diamagnetic term contributes only to the
longitudinal conductivity, since $M_{uv} =0 $ identically when $u\neq v$.
To obtain the frequency response, we Fourier transform Eq. \eqref{eq:chitt}  with respect to the time difference
$\tau = t - t'$,
\beq
\chi_{uv}(\omega,t) =  \int_{\tau=-\infty}^{\tau=0} d\tau~~\chi_{uv}(t,t+\tau) e^{-i\om 
\tau}.\label{eq:lin6}
\eeq
In general, this depends on the probe time $t$. This is especially important
when the perturbation breaks time-periodicity as in the case of a pump-probe experiment and
the results are sensitive to the time of measurement. Additionally, we average this over one
cycle around $t$ to take into account the small but finite width of the probe, 
\begin{small}
\beq
\bar{\chi}_{uv}(\omega,t) = \frac{1}{T}\int_{t}^{t+T} dt' \chi_{uv}(\omega,t'). \label{eq:chiavg}
\eeq
The electrical conductivity is then expressed as
\beq
\si_{uv}(\om) = {\bar{\chi}_{uv}(\om)}/{\om} \label{eq:sigw}.
\eeq
\end{small}
\textit{Response of a system at equilibrium - } For the special case of an unperturbed system,
by noting that stationary states evolve as $|\psi_\al(t)\ra = e^{-i E_\al t}| \psi_\al(0)\ra$
with $E_\al$ being the energy of the $\al$th band, Eq. \eqref{eq:lin6} becomes,
\begin{small}
\bea
\chi^{Eq}_{uv}(\om) =  i \sum_{\al  \ga \vk} &g_{0\al}&\Big[\frac{\la\psi_\al|H_u|\psi_\ga\ra 
\la\psi_\ga|H_v|\psi_\al\ra} {\om + (E_\al  - E_\ga) + i \eta} \non \\ 
&-& \frac{\la\psi_\ga|H_u|\psi_\al\ra
\la\psi_\al|H_v|\psi_\ga\ra} {\om - (E_\al - E_\ga) + i \eta} \Big] \non \\
+ \sum_{\al \vk} & g_{0\al} & \la\psi_\al|M_{uv}|\psi_\al\ra. \label{eq:chieq}
\eea
\end{small}
Note that in the absence of a drive, there is no dependence on the final time $t$ 
as the system is actually time independent and the averaging in Eq. \eqref{eq:chiavg} is
skipped.

\textit{Response of a perfect Floquet drive -}
Similarly, we derive a simpler expression for a Floquet system by noting that the Floquet states
can be written in terms of the Fourier components $|\phi_\al^{(n)}\rangle$ i.e.
\begin{small}
\begin{eqnarray}
|\Psi_\al(t)\rangle = e^{-i\epsilon_\al t}|\phi_\al(t)\rangle = \sum_n e^{-i(\epsilon_\al - \Omega n) t}|\phi_\al^{(n)}\rangle \non \\
|\phi_\al^{(n)}\rangle = \frac{1}{T} \int_0^T dt~e^{-i n \Om t}|\phi_\al(t)\rangle. \label{eq:fftsideband}
\end{eqnarray}
\end{small}
The details of calculating these quasi-mode wavefunctions are given
in the Supplemental Material \cite{supp}.
% \beq
% |\phi_\al(t)\rangle = \sum_{n} e^{i n \Om t}~|\phi_\al^{(n)}\rangle.
% \eeq
The expression for the homodyne \cite{kum1} susceptibility is then modified to,
\begin{small}
\bea
\chi^{Fl}_{uv}(\om)&=&i\sum_{\al\ga m\vk}g_{0\al}\Bigg[\frac{\sum_l 
\langle\phi_\al^{(l)}|H_u|\phi_\ga^{(l+m)}\rangle{\sum_{l'}
	\langle\phi_\ga^{(l'+m)}|H_v|\phi_\al^{(l')}\rangle}} {\om 
+ (\eps_\al - \eps_\ga - m\Om)+i \eta} \non \\
&-&\frac{\sum_l	\langle\phi_\al^{(l)}|H_v|\phi_\ga^{(l+m)}\rangle{\sum_{l' }
		\langle\phi_\ga^{(l'+m)}|H_u|\phi_\al^{(l')}\rangle}} {\om - (\eps_\al - \eps_\ga + 
		m\Om)+i \eta}  \Bigg] \non \\
  &+& \sum_{\al \vk l}g_{0\al}\la \phi_{\al}^{(l)}|M_{uv}|\phi_{\al}^{(l)}\ra. \label{eq:chi_floq}
\eea
\end{small}
The terms in the above expression correspond to optical transitions from the $\alpha$ band of
the $l$th Floquet zone to the $\gamma$ band of the $(l+m)$th Floquet zone. %\sout{While in principle
%the sums over these Fourier indices should be infinite, in practice only a few Fourier components of
%each state are significant.}  
While in principle the Fourier indices being summed over
should span all integers from $-\infty$ to $+\infty$, in practice only a few Fourier components of
each state are significant.
This can be seen by solving a simpler case of a driven single band
system where the weight of the $n$th Fourier mode is proportional to the Bessel function
$\mathcal{J}_n\left(\frac{V_0}{\Omega}\right)$ \cite{Aaron1}. For a small ratio $V_0/\Omega$
%the drive amplitude to its frequency 
this drops rapidly with $|n|$. In our case we find that
$|\phi_\al^{(n)}\ra$ is negligible beyond $n=\pm3$ for the drive parameters that we are
working with. Therefore, in order to numerically evaluate the conductivity, we terminate the
sums in Eq. \eqref{eq:chi_floq} at $l=\pm3$ and $m=\pm3$. For lower drive frequencies or higher
amplitudes a higher cut off may be required.

Importantly, since we assume a perfectly periodic drive we take the population of the levels
to have the simple form, \vspace{-0.3cm}
\begin{small}
\bea
g_{0\al} = \begin{cases} 1 &\mbox{for the lower band~~} (\al=-)\\
	0& \mbox{for the upper band~~} (\al=+).\end{cases}   \label{eq:g_al}
\eea
\end{small}
While this is never the case for a driven system, since all drives are turned on at some finite time,
many authors resort to this population as it is the simplest.
\begin{figure}[t]
\ig[width=8.66cm]{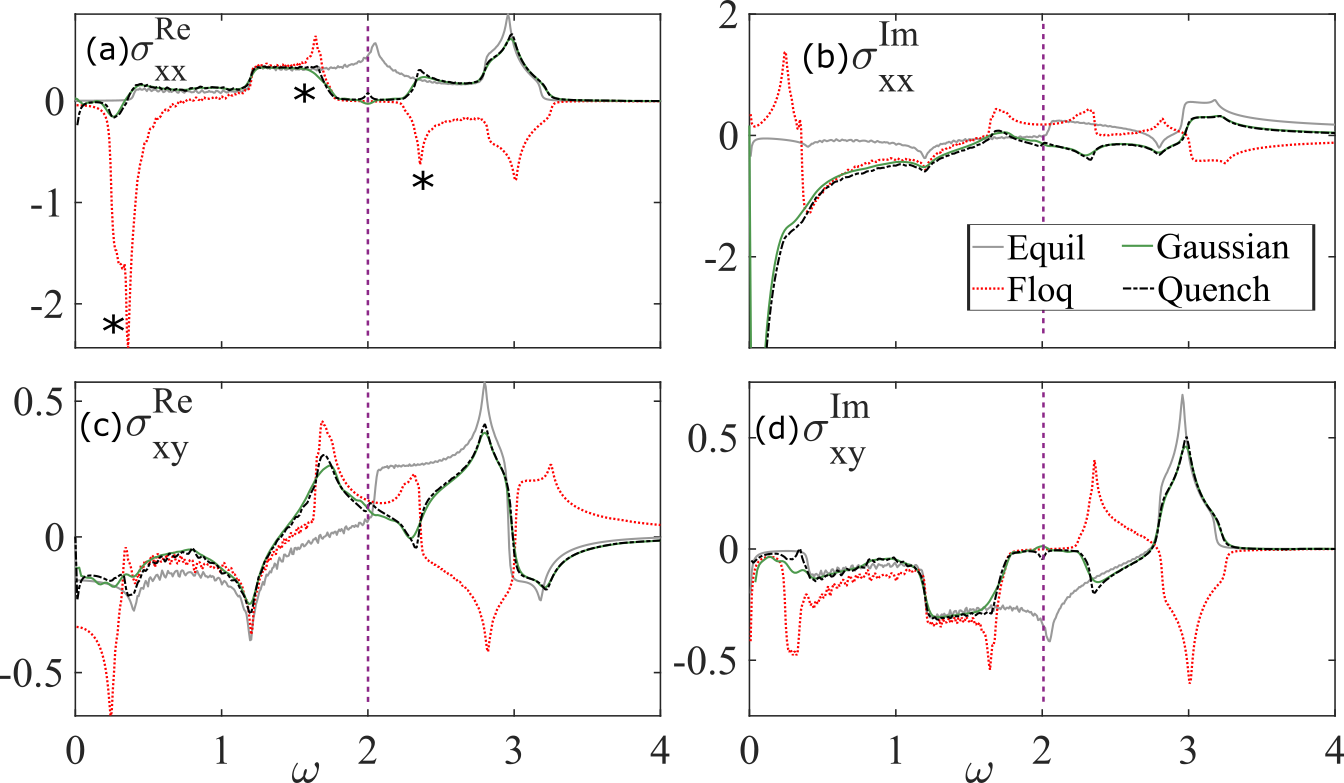}
\caption{Conductivities for the undriven case (grey solid), pure Floquet drive (red dotted), 
a Gaussian pump of width $\Delta= 20$ cycles (green solid) and a
slow quench (black dash-dot) with $\beta = 0.05$. As the response to a Gaussian and quench is almost
identical, it is safe to infer that the response is almost entirely dependent on the instantaneous
drive amplitude.}\label{fig:pk}
\end{figure}

{\it Gaussian and quench pumps - }
We now turn our attention to a realistic scenario where the drive is a Gaussian pulse.
We first compute the conductivity from Eq. \eqref{eq:chitt}
using the real time evolution for a Gaussian pump as well as a quench. We compare that to the 
response of a perfectly periodic drive as well as the unperturbed (equilibrium) case which are obtained
from Eq. \eqref{eq:chi_floq} and Eq. \eqref{eq:chieq} respectively. For the Floquet response we have
used the form of $g_{0\al}$ given in Eq. \eqref{eq:g_al}.

This comparison is shown in Fig. \ref{fig:pk} for $\Om = 2A$ for the real and imaginary parts of
longitudinal and transverse conductivity, i.e. (a) $\si^{Re}_{xx}$, (b) $\si^{Im}_{xx}$,
(c) $\si^{Re}_{xy}$ and (d) $\si^{Im}_{xy}$. The width of the Gaussian is $\Delta = 20$ cycles and the
quench ramp time is $1/\beta = 20$ cycles. The conductivities are shown at the peak of the Gaussian 
(green solid line) and after the quench has reached saturation (black dot-dash line). Although some
features seem to agree, there is a significant difference between the ideal
Floquet response and the actual response with a Gaussian drive or a quench. This difference is more
pronounced when the probe frequency is higher than the drive frequency, i.e., $\om>\Om$ where the sign
of certain features is inverted. Moreover, we see that some features which are very strong in the
Floquet response are suppressed in the Gaussian/quench response. 

\begin{figure}[htb]
\ig[width=8.5cm]{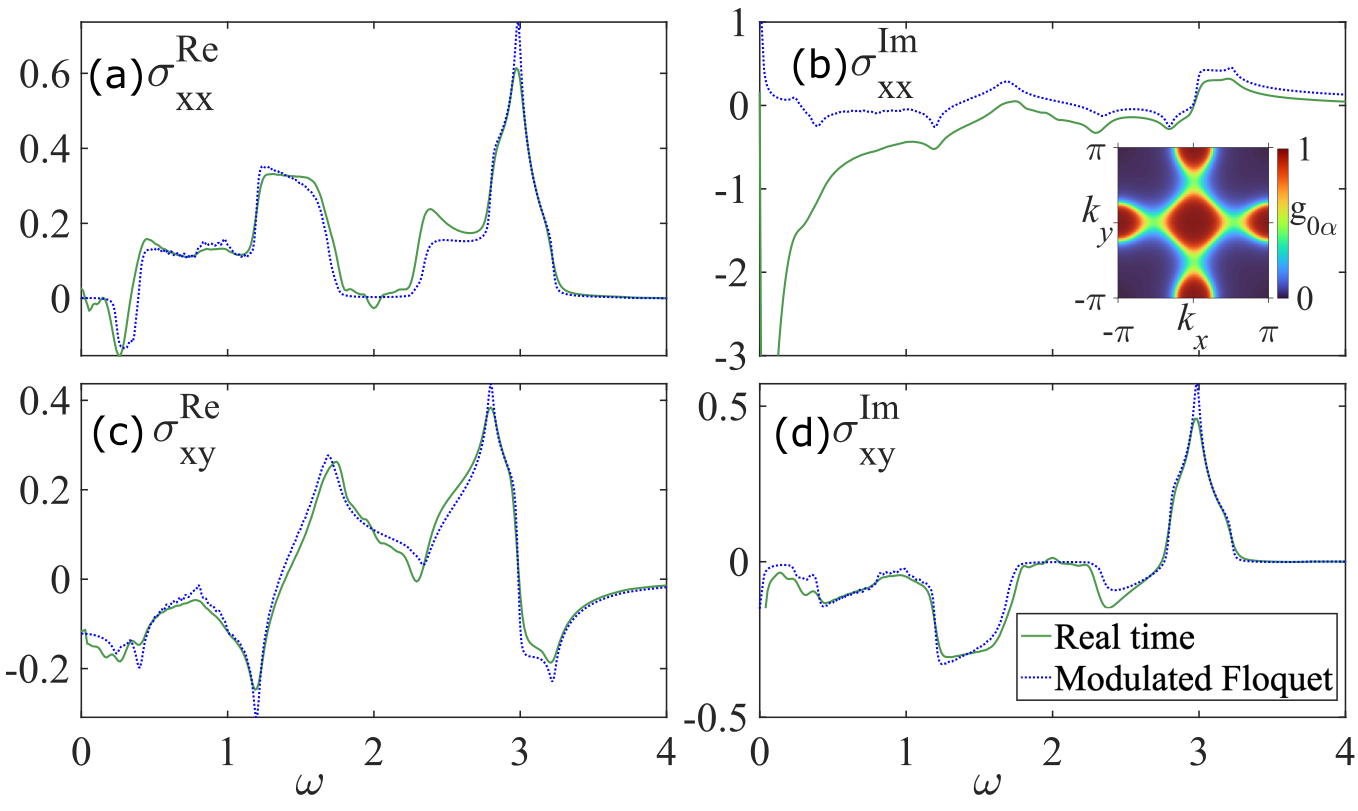}
\caption{Comparison of the response of a Gaussian evaluated using a real time evolution (green) and the
modulated Floquet response (blue) using the $g_{0\alpha}$ in Eq. \eqref{eq:gal_mod} at the peak of
the Gaussian pictured in Fig. \ref{fig:pump}, when the response is expected to have the most similarity
to the ideal Floquet case.}
\label{fig:comp_peak}
\end{figure}

{\it Memory of initial state} - The comparison between the response of the periodically driven systems
and the ones with a pulse shape leads us to speculate that the initial state is not forgotten even
after several cycles of the drive.  To illustrate this we devise an approximate expression
for the time-evolved conductivity as follows.  For a measurement of the Gaussian-driven system at
time $\tilde{t}$ we calculate the Floquet states of a system driven by an ideal sinusoidal
drive whose amplitude is
$V_{{Pump}}(\tilde{t})$.  We then use these states to calculate the response using Eq.~\eqref{eq:chi_floq},
albeit with one important difference.  We replace the simple population $g_{0\al}$ of Eq.~\eqref{eq:g_al} by the
overlap of the Floquet state with the equilibrium state,% \vspace{-0.2cm}
\beq
g_{0\al}(\tilde t)  = |\la\phi_{\al}^{\tilde t}|\Psi_0\ra|^2 \label{eq:gal_mod}
\eeq
where $|\Psi_0\ra$ is the initial state, which in our case we have taken to be lying in the lower band
of the equilibrium spectrum. The Floquet state $|\phi_{\al}^{\tilde t}\ra$ is the eigenstate of the Floquet
operator corresponding to the drive frequency and the instantaneous amplitude, i.e. the magnitude of the
envelope function at the probe time $\tilde t$. We call the response thus obtained as
the ``Modulated Floquet response". The inset in Fig. \ref{fig:comp_peak} shows $g_{0\al}$
at the peak, with the colors showing the population of the lower band of a given Floquet zone. Here
red represents the regions where the lower band is populated and the response in these regions is very close
to the Floquet-like response. On the other hand blue is where the upper band is populated and the response
is exactly inverted from the Floquet response.  Moreover, at the momenta where the original bands have
folded to create the Floquet bands, both bands are partially populated.  This intermediate regime
is responsible for the behaviour around the drive frequency and optical transitions around there are
suppressed. This is marked with an asterisk (*) in Figs. \ref{fig:pk} and \ref{fig:comp_peak}
and corresponds to the transitions shown in the Floquet spectrum of Fig. \ref{fig:pump}(b).

We compare the above ``modulated Floquet response" to that of the time evolved system
for the case of a Gaussian drive at various probe times (See Supplemental material \cite{supp}).
The response well before the peak/ramp is found to be close to the equilibrium response,
as naturally expected. However, even at the peak of the Gaussian pump $(t=0)$, i.e. where
the drive amplitude is the highest, where there is significant mismatch between the 
time-evolved response and either the Floquet or equilibrium response, the modulated
Floquet response gives a good fit. The same holds for the quench scenario, where even
though the drive amplitude is kept on for a significantly long time, the response saturates
and does not replicate the case of a pure Floquet drive. This means that even when an external
driving is switched on very slowly, the system never forgets its initial state and never goes
into a pure Floquet regime where only one of the Floquet bands is fully populated. 

\begin{figure}[htb]
\ig[width=8.5cm]{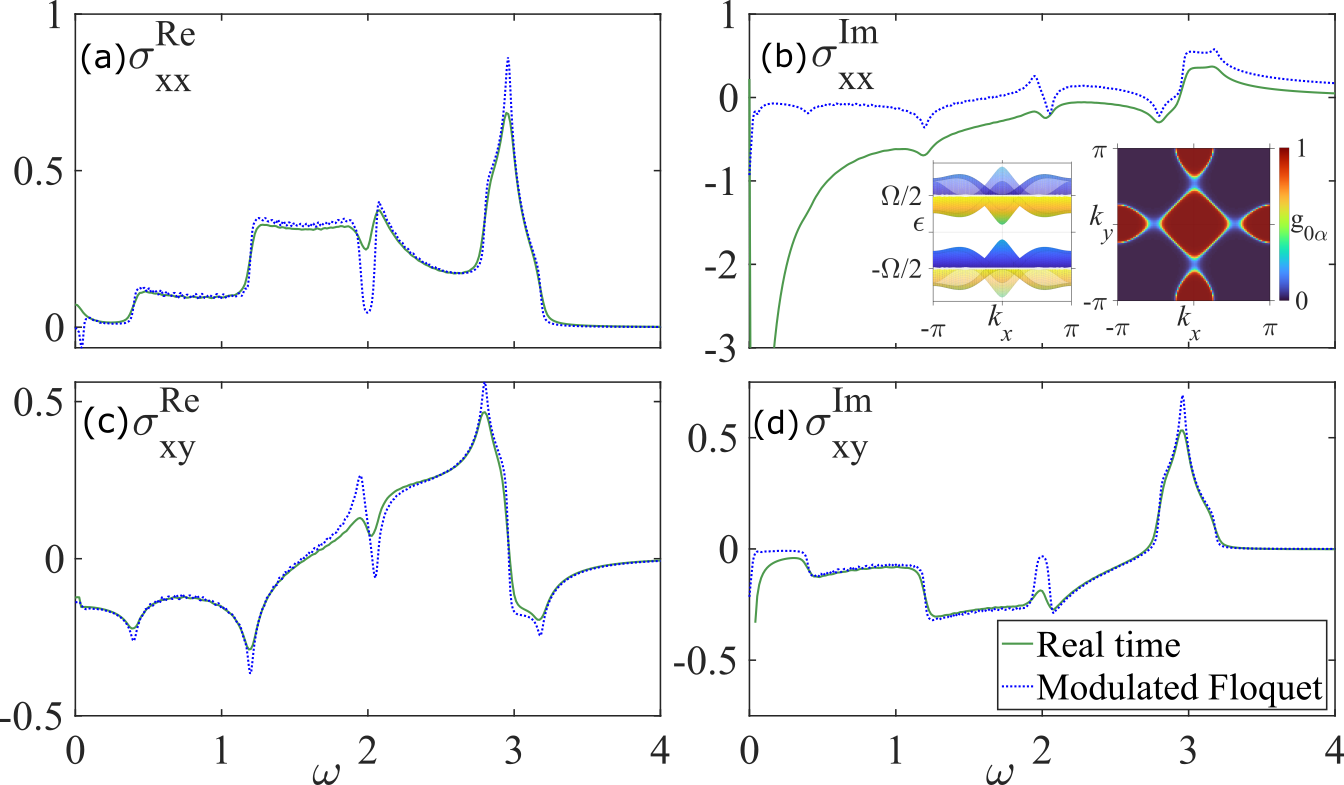}
\caption{Comparison of the response of a Gaussian evaluated using a real time evolution (green) and the
modulated Floquet response (blue) using
the $g_{0\alpha}$ in Eq. \eqref{eq:gal_mod} at $t = -2\Delta$, i.e. well before the peak of the gaussian.}
\label{fig:comp_2Del}
\end{figure}

We also plot  the response obtained from the real time evolution and the
modulated Floquet response for probe time $t=-2\Delta$ in Fig. \ref{fig:comp_2Del}, 
where, again the two behaviours are in agreement. Similar plots for intermediate probe times and for
a higher drive frequency are shown in the Supplemental material \cite{supp}.

{\it Conclusions -} The agreement between the real time evolved conductivity and the modulated Floquet
response is a clear sign of the importance of the initial state at any time of probe, even at the center
of a wide Gaussian shaped pulse or late after the ramp time of a quench.  While the quasi-modes which
contribute to the relevant optical transitions can be approximated by an instantaneous Floquet theory,
one must keep in mind that band inversion may invert energies but does not invert the population of the
bands. This unfortunately means that, unlike at equilibrium, a situation in which quasienergy bands with
topological character are not, in general completely full or empty and quantized DC conductivity is not
likely.   

{\it Acknowledgements - }
The authors thank Babak Seradjeh and Martin Rodriguez-Vega for useful discussions. 
We acknowledge financial support from the Natural Sciences and Engineering Research Council
of Canada (NSERC) and Fonds de recherche du Qu\'ebec – Nature et technologies (FRQNT). RS
acknowledges Dganit Meidan for financial support from Israel Science Foundation (ISF) grant.
The computations presented here were conducted in the computing resources
provided by the Digital Research Alliance of Canada and Calcul Quebec.

\bibliography{refs}

\end{document}

\title{Supplemental Material for \\
Optical conductivity of a topological system driven using a realistic pulse}
\author{Ranjani Seshadri${}^{1,2}$}
\author{T. Pereg-Barnea${}^{2}$}
\affiliation{$^1$Department of Physics, Ben-Gurion University of the Negev,
Beer-Sheva 84105, Israel}
\affiliation{{$^2$Department of Physics, McGill University, Montr\'eal, Qu\'ebec H3A 2T8, Canada}}
\maketitle
\section{Floquet Theory and Side Bands} \label{sup:Floquet}
Consider a time-dependent periodic Hamiltonian $H(t) = H(t+T)$ where $T=2\pi/\Omega$. The  
time-dependent Schrodinger equation (setting $\hbar=1$) can be 	written as
\beq
\Big( H(t)-i \frac{\partial}{\partial t} \Big )|\Psi(t)\ra = 0.\label{eq:TdepSch}
\eeq
According to Floquet theorem \cite{holt, lind, top18a}, the solutions to \eqref{eq:TdepSch} are of the form
$|\Psi(t)\ra  = e^{-i \epsilon_\al t} |\phi_\al (t)\ra$
where $\epsilon_\al$ is the quasienergy and the state $|\phi_\al(t)\ra$ is periodic with the same 
time-period as the hamiltonian $H(t)$, i.e,
\beq
|\phi_\al(t)\ra = |\phi_\al(t+T)\ra.  \label{eq:ftheory}
\eeq
Let us define the Floquet operator, i.e., the time-evolution operator over one drive cycle as,
\beq
\mathcal{U}_T = \mathcal{T} e^{-i \int_{0}^T dt'  H(t')}, \label{eq:FlOp}
\eeq
where $\mathcal{T}$ denotes time ordering.
Since by definition, $|\phi_\al(t+T) \ra= \mathcal{U}_T |\phi_\al(t)\ra$, from Eq. \eqref{eq:ftheory},
\beq
\mathcal{U}_T |\Psi_\alpha (t)\ra = e^{-i \epsilon_\al T}  |\Psi_\alpha(t)\ra, \label{eq:diagU}
\eeq
i.e. , $|\Psi_\alpha(t)\ra$ is an eigenstate of the time evolution operator defined in \eqref{eq:FlOp} with eigenvalue 
$e^{-i \epsilon_\al T}$.  We can write this Floquet state using the Fourier components of $\phi_\al(t)$:
\beq
\Psi_\al(t) = e^{-i\epsilon_\al t}\sum_{n=-\infty}^{\infty} e^{i\Omega n t} \phi^{(n)}_\al =
\sum_{n=-\infty}^{\infty}e^{-i(\epsilon_\al-n\Omega)t} \phi^{(n)}_\al
\eeq
such that $\Psi_\alpha (t)$ is a linear combination of stationary states
$\phi_\alpha^{(n)}(t) = e^{-i(\epsilon_\al-n\Omega)t} \phi^{(n)}_\al$ with energies $\epsilon_\al+n\Omega$. 
The quasienergy $\epsilon_\al$ is therefore defined modulo $\Omega$. We refer to the $n=0$ energy range as the
"Primary Floquet zone" while $n\neq0$ are "side bands".

Clearly, $|\phi(t)\ra$ can be obtained by constructing and diagonalizing the Floquet operator in
Eq. \eqref{eq:FlOp} and the Fourier modes can me obtained by time evolving $\Psi(t)$ over a single drive cycle.

%\beq
%|\phi(t)\ra = \mathcal{U}_{t} |\phi(0)\ra 
%\eeq
%where $\mathcal{U}_t = \mathcal{T} e^{-i \int_{0}^t dt'  H(t')}$ for any general time t. The side bands can then
%be written as the Fourier components of these states modulo the micromotion
%\beq
%|\phi_\al^{(n)}\rangle = \frac{1}{T} \int_0^T dt~e^{-i n \Om t}~e^{i\eps_\al t}|\phi_\al(t)\rangle.
%\eeq

%--------------------------------------------------------------------------------------------------------~%
\section{Evaluating Longitudinal and Transverse conductance using Linear response theory} \label{app:linresp}
Since our system does not obey time-translation invariance we wish to write the response functions in real
time as a function of  both the pump time (measured from some initial time $t_0$ and the probe time.  

According to the Kubo formula the susceptibility is given by,
\bea
\chi_{AB}(t,t')=\lim_{\eta\rightarrow 0^+} e^{\eta t'}\Bigg(i\Theta(t-t')Tr\Big\{g_0 [A^I(t'),B^I(t)]\Big\}
+ \delta(t-t')Tr \Big\{g_0 M^I(t)\Big\}\Bigg) \label{eq:lin1gen}
\eea
Here the operators $A^I$, $B^I$ and $M^I$ are in the interaction representation.

Let us assume that the system is in the initial state $|\Psi_{0}\ra \equiv |\Psi_{0}(t_0)\ra$. Since we wish
to study the conductivity - both longitudinal and transverse - we are interested in the current operators in
the $x$ and $y$ directions. The current operator in the $u-$direction in the interaction representation is
written as 
\beq
J^I_{q,u}(t) = U(t,t_0) J_{q,u}(t) U^\dag(t,t_0), \label{eq:defJ}
\eeq
with the time evolution operator being defined as
\beq
U(t,t_0) = \sum_\lam |\Psi_\lam(t)\ra \la \Psi_\lam(t_0)|, \label{eq:defU}
\eeq
where summation is over all the bands. The paramagnetic term can therefore be written as
\bea
Tr\Big\{g_0 [J^I_{q,u}(t'),J^I_{-q,v}(t)]\Big\} &=& \la \Psi_0 | U(t,t_0) J_{q,u}(t) 
U (t_0,t) U(t',t_0) J_{-q,v}(t') U (t_0,t') |\Psi_0\ra- \{u \leftrightarrow v\} ~~~~~~~~~~\non \\
&=&\la \Psi_0(t)| J_{q,u}(t) U (t_0,t) U(t',t_0) J_{-q,v}(t')|\Psi_0(t')\ra -
\{u \leftrightarrow v\} ~~~~~~~~~~~\non\\
%\text{Inserting a complete set of states at $t_0$, this becomes}\non \\
&=&\la \Psi_0(t)| J_{q,u}(t) U (t_0,t) \Big(\sum_\lam |\Psi_\lam(t_0)\ra \la \Psi_\lam(t_0)
|\Big) U(t',t_0)  J_{q,v}(t')|\Psi_0(t')\ra - \{u \leftrightarrow v\}
\eea
where in the last step we have used  
$\sum_\lam |\Psi_\lam(t_0)\ra \la \Psi_\lam(t_0)|= \mathbb{1}$
Therefore,
\beq
Tr\Big\{g_0 [J^I_{q,u}(t'),J^I_{-q,v}(t)]\Big\} = \sum_\lam \la \Psi_0(t) J_{q,u}(t)
\Psi_\lam(t)\ra \la \Psi_\lam(t') J_{-q,v}(t) \Psi_0(t')\ra -\{u\leftrightarrow v\}
\eeq

Now the current operator is defined as
\beq
J_{q,u}(t) = \sum_{p,\al,\beta} \frac{\pa h^{\al\be}(t)}{\pa p_u} c^\dag_{p+\frac{q}{2},\al} c_{p-\frac{q}{2},\be}
\eeq
The many body state for a non-interacting system is a product over all momenta, i.e.,
\beq
|\Psi_\lam(t)\ra = \prod_{k,\si} \Psi^\lam_{k,\si}(t) c^\dag_{k,\si} |0\ra
\eeq

Taking the limit $q\rightarrow 0$ The general expression for conductance in Eq. \eqref{eq:lin1gen} can be
simplified to
   
\bea
\chi_{uv}(t,t') =  \lim_{\eta\rightarrow 0^+} e^{\eta t'} \sum_{\al \ga \vk} g_{0\al} (\vk) \Big[2 i \Theta(t-t')
&Im& \Big(\la\Psi_\al(\vk,t')|\frac{\partial{H(\vk)}}{\partial{k_u}}|\Psi_\ga(\vk,t')\ra \la\Psi_\ga(\vk, t)
|\frac{\partial{H(\vk)}}{\partial{k_v}}|\Psi_\al(\vk,t)\ra \Big) \non \\
&+& \delta(t-t')\la\Psi_\al(\vk,t)|\frac{\partial^2H(\vk)}{\partial_{k_u}\partial_{k_v}}
|\Psi_\al(\vk,t)\ra\Big]\label{eq:chitt}
\eea

where the summation is over all degrees of freedom: bands as well as momentum. Assuming that the system is in its ground
state to begin with and only the lowest band is populated, the density matrix for a two-band model (such as BHZ model)is
given by,
\bea
g_{0\al} = \begin{cases} 1 &\mbox{for lower band } \\
	0& \mbox{for upper band}.\end{cases}.   \label{eq:lin3}
\eea
Other initial conditions can be easily taken into account by  changing $g_{0\al}$.

It should be noted that the above two-times susceptibility reduces to the well known equilibrium result
(Eq. (8) in the main text) when we assume time translation invariance and to the Floquet result (Eq. (10) in the
main text) when we assume a perfectly periodic Hamiltonian without a pulse/turn-on envelope.
%----------------------------------------------------------------------------------------------------------------~%
\section{Results for drive frequency $\Om = 2A$} \label{sec:om2}
\begin{figure}[H]
\centering
\ig[width=17.9cm]{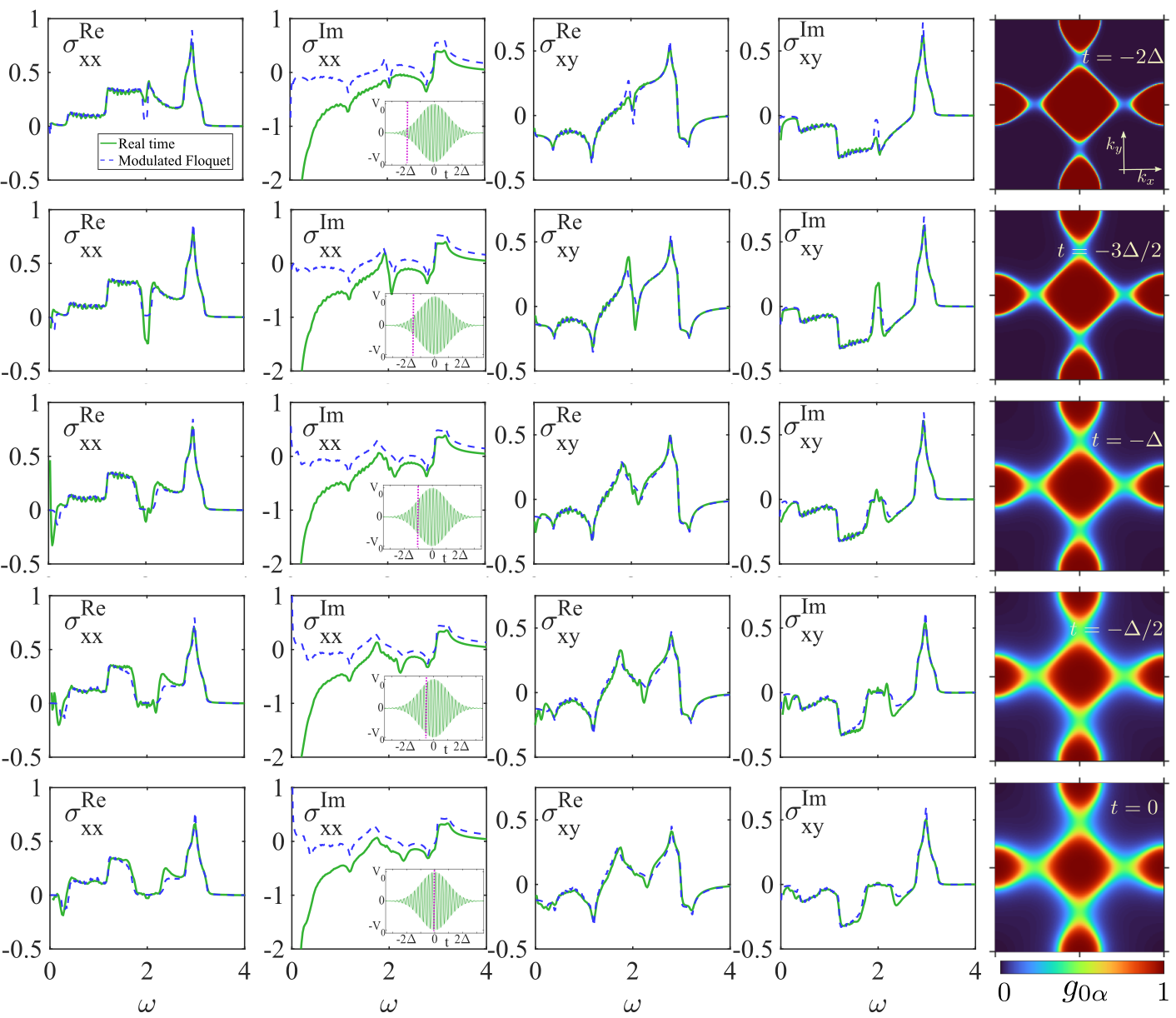}
\caption{Conductivities and population $g_{0\al}$ for probe times $t = -2\De$, $-3\De/2$, $-\De$, $-\De/2$, and $0$
(top to bottom) where $t=0$ represents the peak of the Gaussian as shown in Fig. 1 of the main text.
The probe time is indicated by a vertical magenta line in the inset of
each of the rows. The green (solid) and the blue (dashed) lines in each of the panels shows the conductivities obtained
from the real time simulation and the Modulated Floquet simulations respectively. The colorbar for $g_{0\alpha}$
is in the bottom right, with the blue (red) representing the top (bottom) band being fully populated. As the peak
of the Gaussian approaches, the region of the B.Z. where both the bands have significant population increases, and 
hence the feature close the drive frequency ($\Om=2A$ in this case) becomes broader and broader. This can be seen in
all four components of conductivity.}
\label{fig:W2timelapse}
\end{figure}

In the main text (Figs. 3 and 4), we have shown the comparison between the conductivities obtained by the real time
and modulated Floquet simulations and how these relate to the population of the Floquet bands $g_{0\alpha}$. Here in
Fig. \ref{fig:W2timelapse} we have shown these results for various probe times $t = -2\De$, $-3\De/2$, $-\De$, $-\De/2$,
and $0$. The probe/measurement time with respect to the peak of the Gaussian is shown as a vertical magenta line in the
inset of each of the rows. The green (solid) and the blue (dashed) lines in each of the panels shows the conductivities
obtained from the real time simulation and the Modulated Floquet simulations respectively. The rightmost surface plot in
each row shows the instantaneous $g_{0\alpha}$ for the bottom band in one Brillouin zone.  It is defined as the overlap
of the lower Floquet band, calculated with the drive amplitude at time $t$, with the initial equilibrium state. Blue
(red) represents the region where the upper (lower) band is almost entirely populated as shown in the colorbar in the
bottom right. As the peak of the Gaussian approaches, the region of the B.Z. where both the bands have significant
population (green-yellow on the color scale) increases, and hence the feature close to the drive frequency
($\om = \Om=2A$ in this case) becomes broader and broader. Similarly the feature close to $\om=0$ becomes more
pronounced as this represents the transition from one Floquet zone to the next. This can be seen in all four
components of conductivity.

%----------------------------------------------------------------------------------------------------------------~%

\section{Results for higher drive frequency $\Om =3$}
\begin{figure}[htb]
\centering
\ig[width=9cm]{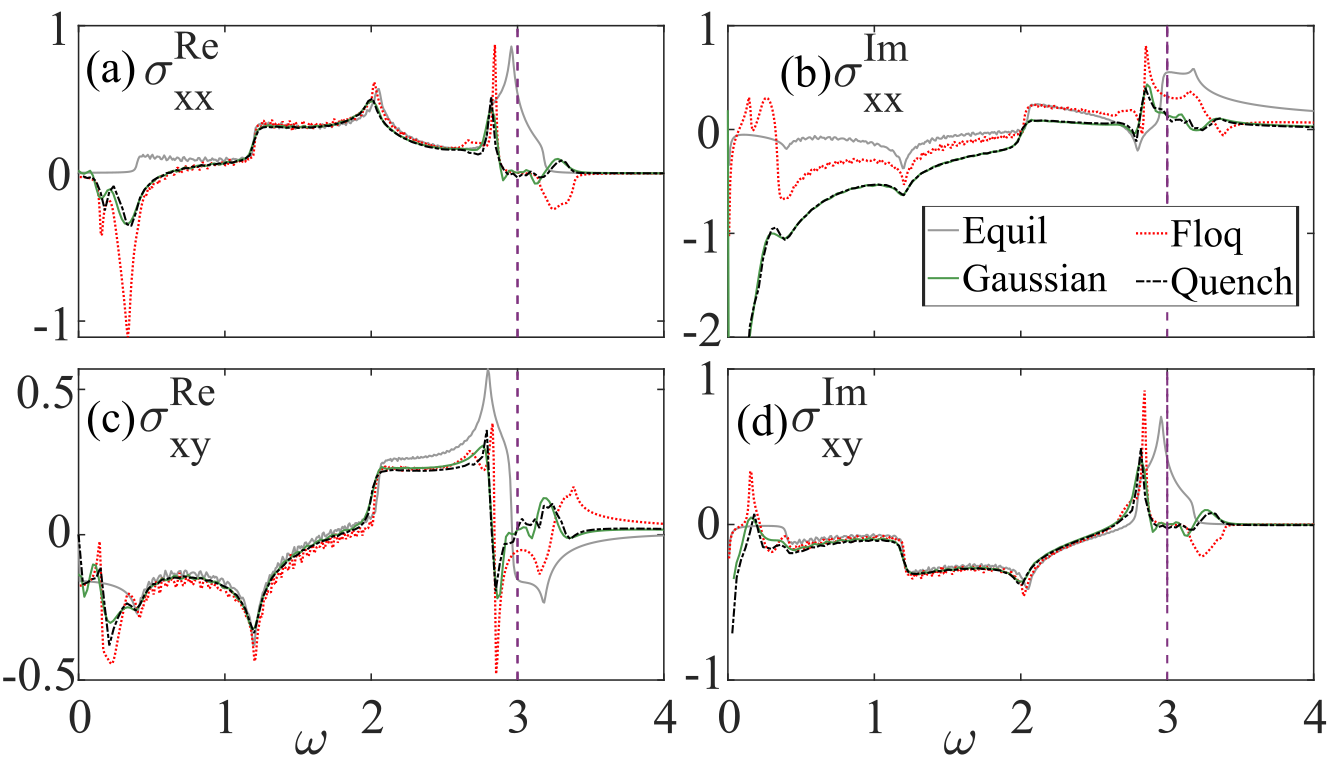}
\ig[width=4.5cm]{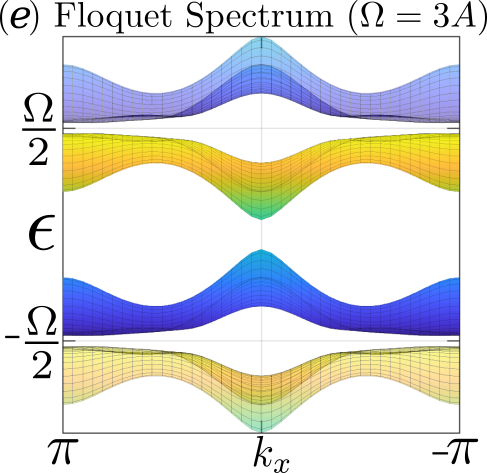}
\caption{Comparison of the conductance ((a)-(d)) of a BHZ system in the undriven/equilibrium case
(grey solid),
pure Floquet drive (red dotted), a Gaussian pump with a width $\Delta= 20$ cycles (green solid) and a
slow quench (black dash-dot) with $\beta = 0.05$. Since the response to a Gaussian and Quench is almost
identical, it is safe to infer that the response is almost entirely dependent on the instantaneous
amplitude of the drive. The drive frequency $\Om= 3.0A$. The vertical line marks where the probe frequency
$\om = \Om$. (e) shows the Floquet spectrum for $\Omega = 3A$ and $V_0 = 0.35 A$}\label{fig:pk_spp}
\end{figure}

Here we show the response of the BHZ system to a higher driver frequency, following the same procedure
as discussed in the main text. First, in Fig. \ref{fig:pk_spp} we compare the response when driven using a
Gaussian and Quench (green and black respectively) to the equilibrium and Floquet case. Again, as observed in
the main text for $\Om=2A$, the behavior with a Gaussian/Quench in the regime $\om<\Om$ resembles the
Floquet response, while that in the regime $\om>\Om$ changes sign.

\begin{figure}[H]
\centering
\ig[width=17.9cm]{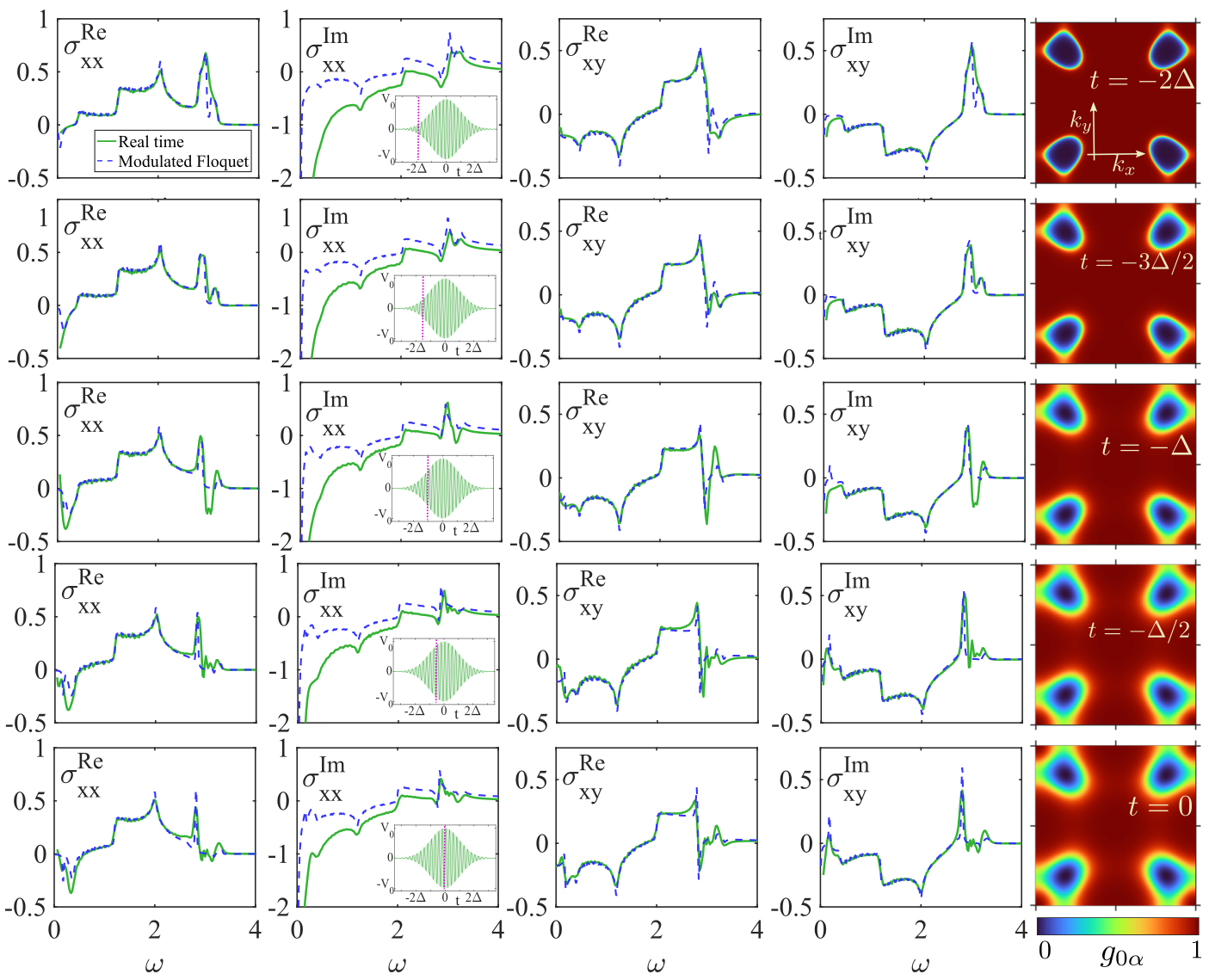}
\caption{Conductivities and population $g_{0\al}$ for various probe times $t = -2\De$, $-3\De/2$, $-\De$, $-\De/2$, and $0$
(top to bottom) where $t=0$ represents the peak of the Gaussian. The drive frequency $\Om=3A$ which is slightly less
than the spectral width. The probe time is indicated by a vertical magenta line in the inset of
each of the rows. The green (solid) and the blue (dashed) lines show the conductivities obtained
from the real time simulation and the Modulated Floquet simulations respectively. The colorbar for $g_{0\alpha}$
in the bottom right, with the blue (red) representing the top (bottom) band being fully populated.}
 \label{fig:comp_peak}
\end{figure}

Similar to Sec. \ref{sec:om2}, we now compare, for $\Om=3A$, the real time and Modulated Floquet response,
with the latter involving the new population of states $g_{0\alpha}$ (as defined by Eq. (12) of the main
text). Again in Fig. \ref{fig:comp_peak}, we have shown the results for various probe times $t = -2\De$, $-3\De/2$, $-\De$, $-\De/2$, and $0$.

The probe/measurement time with respect to the peak of the Gaussian is shown as a vertical magenta line in the
inset of each of the rows. In each of the panels, the green (solid) and the blue (dashed) lines depicts the
conductivity obtained from the real time simulation and the Modulated Floquet simulations respectively. The
rightmost plot in each row is the instantaneous $g_{0\alpha}$ for the bottom band in one Brillouin zone shown
as a surface plot. It is defined as the overlap of the lower Floquet band, calculated with the drive amplitude at
time $t$, with the initial equilibrium state. Blue (red) represents the region where the upper (lower) band is
almost entirely populated as shown in the colorbar in the bottom right. As the peak of the Gaussian approaches,
the region of the B.Z. where both the bands have significant population (green-yellow on the color scale) increases,
and hence the feature close to the drive frequency ($\om = \Om=2A$ in this case) becomes broader and broader.
Similarly the feature close to $\om=0$ becomes more pronounced as this represents the transition from one Floquet
zone to the next.
One stark difference between $g_{0\al}$ for $\Om=2A$ and $\Om = 3A$ is that in the latter case, the area of
the Brillouin zone in which the upper band is significantly populated is much less than in the former case.

%----------------------------------------------------------------------------------------------------------------~%
\section{Numerical evaluation of Chern number}\label{sec:app_num}
Given a momentum-space hamiltonian $h(\vk)$, the eigenvalue equation is given by
\beq
h(\vk) |\Psi_\al(\vk)\ra = E_\al(\vk)|\Psi_\al(\vk)\ra 
\eeq
with $|\Psi_\al(\vk)\ra$ being the eigenstate corresponding to eigenvalue $E_\al$. For any two-band model, such as
the Bernevig-Hughes-Zhang model which we have considered in this work, the index $\al = \pm$ corresponds to the upper/lower
bands and $|\Psi_\al(\vk)\ra$s are two-component spinors. However, the discussion the follows is applicable to any $n-$band
model (with each eigenstate being an $n-$component spinor).

The Berry curvature $C_\al$ of the $\al-$th band can be written in terms of the eigenstate $|\Psi_\al(\vk)\ra$, 
\bea 
C_\al =~ \frac{i}{2\pi} ~\int \int dk_x dk_y \Big[ ~
\frac{\partial \Psi_\al^\dg}{\partial k_x} \frac{\partial \Psi_\al}{\partial k_y} -
\frac{\partial \Psi_\al^\dg}{\partial k_y} \frac{\partial \Psi_\al}{\partial k_x} \Big].
\label{eq:chernapp}
\eea
with $\Psi_\al(\vk) = \la \vk| \Psi_{\al}(\vk)\ra$ being the momentum-state representation of the
eigenstate.

When $\Psi_\al(\vk)$ is numerically evaluated, an arbitrary phase factor is included. Therefore, the direct
numerical evaluation of the Chern number using the expression in Eq. \ref{eq:chernapp} becomes complicated.
In other words, we need a method to calculate $C_\al$ which is gauge-invariant, i.e. independent of the
arbitrary phase factor which can always multiply the eigenvector. Therefore we resort to the method similar
to the one prescribed by Fukui et.al. \cite{fukui}.

Consider an actual numerical computation where the eigenvectors are evaluated on a discretized two-dimensional Brillouin
Zone (B.Z). We denote the points on the momentum mesh as $(n_x,n_y)$ with $n_x\in [1,N_x], n_y \in [1,N_y]$. The $\al$th
eigenvector at each such momentum point is written as $|\Psi_{\al,n_x,n_y}\ra$. Four consecutive $k-$points on this
momentum mesh form a plaquet. 

We define a variable $F_{\al,n_x,n_y}$ associated with this plaquet as
\begin{small}
\beq
F_{\al,n_x,n_y} =\ln \Bigg[\frac{\la\Psi_{\al,n_x,n_y}|\Psi_{\al,n_x+1,n_y}\ra}{|\la\Psi_{\al,n_x,n_y}|
\Psi_{\al,n_x+1,n_y}\ra|} \frac{\la\Psi_{\al,n_x+1,n_y}|\Psi_{\al,n_x+1,n_y+1}\ra}{|\la\Psi_{\al,n_x+1,n_y}|
\Psi_{\al,n_x+1,n_y+1}\ra|} \frac{\la\Psi_{\al,n_x+1,n_y+1} |\Psi_{\al,n_x,n_y+1}\ra}{|\la\Psi_{\al,n_x+1,n_y+1}
|\Psi_{\al,n_x,n_y+1}\ra|} \frac{\la\Psi_{\al,n_x,n_y+1}|\Psi_{\al,n_x,n_y}\ra}{|\la\Psi_{\al,n_x,n_y+1}|
\Psi_{\al,n_x,n_y}\ra|}\Bigg]
\label{eq:Fplaq}.\non \\ \non \\
\eeq
\end{small}
Note that we use only the principal branch of the logarithm function in Eq. \eqref{eq:Fplaq}. Summing this over
the entire Brillouin Zone gives the Chern number of the $\al$th band, which is an integer
\beq
C_{\al} = \frac{1}{2 i \pi}\sum_{n_x,n_y} F_{\al,n_x,n_y}.
\eeq
\bibliography{refs}